%% file: main.tex
\documentclass[aps,pra,twocolumn,superscriptaddress,nofootinbib,floatfix]{revtex4-2}

\usepackage[T1]{fontenc}
\usepackage{mathptmx}
\usepackage{amsmath,amssymb,amsthm}
\usepackage{graphicx}
\usepackage{xcolor}
\usepackage{enumitem}
\usepackage{braket}
\usepackage[colorlinks=true,linkcolor=blue!50!black,citecolor=blue!50!black,urlcolor=blue!50!black]{hyperref}
\usepackage{microtype}

\newtheorem{lemma}{Lemma}
\input{numbers.tex}

\begin{document}

\title{Spacetime Sentinels: a Qiskit game that exposes the competition between spacetime checks and magic in low-overhead quantum error detection}

\author{Sumit Tapas Chongder}
\thanks{These authors contributed equally to this work.}
\email{m25iqt013@iitj.ac.in}
\affiliation{Department of Computer Science \& Engineering and Department of Physics, Indian Institute of Technology Jodhpur, Jodhpur, Rajasthan, India}

\author{Aditya Kumar}
\thanks{These authors contributed equally to this work.}
\email{m25iqt002@iitj.ac.in}
\affiliation{Department of Computer Science \& Engineering and Department of Physics, Indian Institute of Technology Jodhpur, Jodhpur, Rajasthan, India}

\date{October 2026}

\begin{abstract}
We present \emph{Spacetime Sentinels}, a game in which the player protects a noisy circuit with spacetime Pauli checks and is scored on the post-selected fidelity and the fraction of shots that survive, with every number computed in Qiskit. The player spends a budget of ancilla qubits on checks found by the \texttt{qiskit-paulice} add-on, must inject a quota of $T$ gates into the protected circuit, and chooses whether to place them where they commute with every check or to ignore the checks. Each move is evaluated exactly with Qiskit Aer density-matrix simulation against a stabilizer or statevector reference, and the evaluation agrees with 50{,}000-shot runs of the \texttt{SamplerV2} primitive. We score moves with a fidelity rate $R=F_{\mathrm{ps}}\kappa^{1/2}$, whose inverse square is the sampling overhead of detection followed by rescaling. Exhaustive evaluation of all \NumMoves{} legal moves under two noise models, together with sweeps over five random payloads, reveals three trade-offs that are not obvious before playing. Checks pay for themselves only above a break-even depth: at depth 6 adding checks lowers $R$, while at depth 18 the best number of checks raises it by a factor of \EoneGainDeep{} and lifts the fidelity from \EoneFzeroDeep{} to \EoneFpsDeep. $T$ gates add no noise on hardware, yet each check removes spacetime locations where a $T$ gate can be placed safely, from all 90 at depth 18 with no checks to about 14 with four, so the optimal number of checks falls as the magic quota grows. Finally, a check broken by a $T$ gate is worse than no check: with sixteen $T$ gates at depth 18, blind placement lowers the post-selected fidelity to \EtwoBlindFDeepTSixteen, and retiring the broken checks lowers it further to \EtwoRetireFDeepTSixteen, below the \EtwoNoneFDeepTSixteen{} of the unprotected circuit. The game runs in the browser and in the terminal, and all code and data are open source.
\end{abstract}

\maketitle

\input{sections/intro}
\input{sections/background}
\input{sections/game}
\input{sections/methods}
\input{sections/results}
\input{sections/discussion}

\bibliography{refs}

\end{document}

%% file: numbers.tex
\newcommand{\BestDeepTEightGain}{1.135}

\newcommand{\BestDeepTEightTargets}{q0,q3,q4}
\newcommand{\BestDeepTFourFps}{0.910}
\newcommand{\BestDeepTFourFzero}{0.632}

\newcommand{\BestDeepTFourKeep}{0.615}

\newcommand{\BestDeepTFourTargets}{q0,q3,q4}

\newcommand{\BestDeepTSixteenGain}{1.120}

\newcommand{\BestDeepTSixteenTargets}{q2,q3}

\newcommand{\BestShallowTZeroGain}{1.018}

\newcommand{\BestShallowTZeroTargets}{q0,q4}
\newcommand{\BestTableRows}{6 & 0 & $q_{0,4}$, Clifford & 0.828 & 0.914 (0.851) & 1.018 \\ 6 & 1 & $q_{0,4}$, safe & 0.828 & 0.914 (0.851) & 1.018 \\ 6 & 2 & $q_{0,4}$, safe & 0.828 & 0.914 (0.851) & 1.018 \\ 6 & 4 & $q_{0,4}$, safe & 0.832 & 0.914 (0.851) & 1.013 \\ 6 & 8 & $q_{0,4}$, safe & 0.831 & 0.911 (0.851) & 1.011 \\ 6 & 16 & $q_{0,4}$, safe & 0.833 & 0.911 (0.851) & 1.009 \\ 12 & 0 & $q_{0,3}$, Clifford & 0.722 & 0.896 (0.750) & 1.074 \\ 12 & 1 & $q_{0,3}$, safe & 0.722 & 0.896 (0.750) & 1.074 \\ 12 & 2 & $q_{0,3}$, safe & 0.722 & 0.896 (0.750) & 1.074 \\ 12 & 4 & $q_{0,3}$, safe & 0.722 & 0.894 (0.750) & 1.072 \\ 12 & 8 & $q_{0,3}$, safe & 0.722 & 0.894 (0.750) & 1.073 \\ 12 & 16 & $q_{0,3}$, safe & 0.722 & 0.895 (0.750) & 1.074 \\ 18 & 0 & $q_{0,3,5}$, Clifford & 0.637 & 0.910 (0.615) & 1.119 \\ 18 & 1 & $q_{0,3,4}$, safe & 0.637 & 0.909 (0.615) & 1.119 \\ 18 & 2 & $q_{0,3,4}$, safe & 0.634 & 0.910 (0.615) & 1.126 \\ 18 & 4 & $q_{0,3,4}$, safe & 0.632 & 0.910 (0.615) & 1.130 \\ 18 & 8 & $q_{0,3,4}$, safe & 0.629 & 0.910 (0.615) & 1.135 \\ 18 & 16 & $q_{2,3}$, safe & 0.629 & 0.863 (0.666) & 1.120}
\newcommand{\CapDeepFour}{14.3}

\newcommand{\CapDeepSix}{7.4}
\newcommand{\CapDeepTwo}{29.4}

\newcommand{\CapTotalDeep}{90}

\newcommand{\EfiveGainDeepHigh}{1.177}
\newcommand{\EfiveGainDeepLow}{1.009}

\newcommand{\EfiveGainShallowHigh}{0.906}
\newcommand{\EfiveGainShallowLow}{0.997}

\newcommand{\EoneFpsDeep}{0.886}

\newcommand{\EoneFzeroDeep}{0.629}

\newcommand{\EoneGainDeep}{1.103}
\newcommand{\EoneGainMedium}{1.037}

\newcommand{\EoneGainOneShallow}{0.993}

\newcommand{\EoneKDeep}{3}
\newcommand{\EoneKMedium}{3}

\newcommand{\EoneKeepDeep}{0.612}

\newcommand{\EsevenFpsHighTSixteen}{0.837}
\newcommand{\EsevenFpsHighTZero}{0.890}

\newcommand{\EsevenFzeroHighTSixteen}{0.640}
\newcommand{\EsevenFzeroHighTZero}{0.721}

\newcommand{\EtwoBlindFDeepTSixteen}{0.158}

\newcommand{\EtwoBlindKDeepTSixteen}{0.174}

\newcommand{\EtwoBrokenFourDeep}{2.4}

\newcommand{\EtwoNoneFDeepTSixteen}{0.628}

\newcommand{\EtwoNoneFzeroTSixteenDeep}{0.628}

\newcommand{\EtwoNoneFzeroTZeroDeep}{0.629}

\newcommand{\EtwoRetireFDeepTSixteen}{0.102}

\newcommand{\EtwoSafeDrift}{0.004}

\newcommand{\EtwoSafeFDeepTSixteen}{0.892}

\newcommand{\EtwoSafeKDeepTSixteen}{0.604}

\newcommand{\FezGainDeep}{1.060}

\newcommand{\FezGainShallow}{1.015}
\newcommand{\FezKDeep}{3}

\newcommand{\FzeroDeep}{0.637}
\newcommand{\FzeroMedium}{0.722}
\newcommand{\FzeroShallow}{0.828}
\newcommand{\GridMinutes}{53}
\newcommand{\HzeroDeep}{1.000}
\newcommand{\LocMaxDeep}{1.106}

\newcommand{\LocMinDeep}{1.062}

\newcommand{\NumIllegal}{25}
\newcommand{\NumMoves}{5{,}412}

\newcommand{\ProFez}{0.0088}
\newcommand{\PtwoFez}{0.0048}
\newcommand{\ValHdiff}{0.011}
\newcommand{\ValKeepSigma}{1.4}
\newcommand{\ValStabDiff}{3e-15}

%% file: sections/intro.tex
\section{Introduction}
\label{sec:intro}

Students usually meet quantum error correction as a list of facts: a code is a subspace, a stabilizer is a Pauli operator, a syndrome is a bit string, and a decoder turns the bit string into a correction. The facts are correct, but they hide the part that makes fault tolerance hard in practice, namely that every protective measure is itself built out of noisy gates and has to pay for itself. The question posed by the Qiskit Fall Fest 2026 challenge~\cite{qff2026prompt} is how to let a student feel that tension directly. Our answer is a game in which every move is a design decision about a real error-detection scheme and every score is the outcome of a Qiskit simulation of that scheme.

The scheme is low-overhead error detection with spacetime codes~\cite{martiel2025spacetime}, available in Qiskit as the \texttt{qiskit-paulice} add-on~\cite{paulice,ibm_tutorial_spacetime}. A spacetime check is a single ancilla qubit that touches one target qubit of a Clifford circuit at several moments, applying a controlled Pauli each time, and is then measured. Combined with the parity of some output bits, the ancilla outcome is deterministic on an ideal run. A non-trivial outcome therefore flags an error, and the shot is discarded. The construction descends from coherent Pauli checks~\cite{vandenberg2023cpc} and coherent parity check codes~\cite{roffe2018cpc,chancellor2023graphical,eczoo_cpc}, but it spreads each check over time instead of measuring a heavy operator at one instant, which keeps it compatible with sparse hardware connectivity. IBM has used the same idea to sample from a 64-qubit, depth-73 Clifford circuit doped with 314 $T$ gates while certifying a lower bound on the fidelity~\cite{martiel2026sampling,ibm_blog_trusted}.

Where does such a scheme sit? Following the distinction drawn in the challenge text and in Ref.~\cite{ibm_blog_emc}, \emph{error detection} flags the errors its checks can see and throws those shots away. It does not repair the state, and errors that the checks cannot see go through untouched. \emph{Error mitigation} leaves the hardware as it is and corrects the estimate of an observable in post-processing, at the cost of extra circuits and shots~\cite{temme2017error}. \emph{Error correction} encodes logical qubits in many physical qubits, measures stabilizers repeatedly and uses a decoder to undo errors while the computation runs~\cite{shor1995scheme,fowler2012surface,krinner2022realizing,google2025below}. Our game lives entirely in the first category. The player never corrects anything; the only currency is the fraction of shots that survive post-selection.

\paragraph*{Design goal.} The challenge asks for a game whose rules correspond to the decisions of the error-detection workflow and whose score comes from computed numbers, not from hand-set values. We add two requirements of our own. First, the game should contain at least one trade-off that a strong student would not guess before playing. Second, the numbers behind the game should be good enough to stand as a small research study: exact where possible, validated against shot-based sampling, and averaged over several random circuits.

\paragraph*{Contributions.}
\begin{enumerate}[leftmargin=*,itemsep=2pt]
  \item \textbf{A game whose rules are the physics.} \emph{Spacetime Sentinels} asks the player to spend an ancilla budget on spacetime checks, to inject a quota of $T$ gates, and to choose how the $T$ gates are placed relative to the checks. The rule-to-physics map is given in Table~\ref{tab:map}. The game runs in the browser and in the terminal.
  \item \textbf{Every score from Qiskit.} Checks are found with \texttt{add\_pauli\_checks}, ideal states come from stabilizer or statevector simulation, and noisy post-selected states come from exact Qiskit Aer density-matrix simulation. We evaluated all \NumMoves{} legal moves of the game under two noise models, one taken from the challenge text and one derived from the calibration data of a Heron backend.
  \item \textbf{A principled score.} We combine post-selected fidelity and kept fraction into a fidelity rate $R_\beta=F_{\mathrm{ps}}\,\kappa^{\beta}$. For $\beta=1/2$ this is the signal-to-noise ratio per square root of a raw shot, and $R_{1/2}^{-2}$ is the sampling overhead of detection followed by rescaling.
  \item \textbf{Three non-obvious findings, backed by numbers.} (i) Checks do not pay for themselves on shallow circuits; there is a break-even depth. (ii) $T$ gates are free on the hardware but expensive for the code: every check removes the spacetime locations where a $T$ gate can be placed without breaking a check, so protection and magic compete for the same wires. (iii) A check broken by a $T$ gate is worse than no check at all, and it keeps hurting even after the player stops listening to it.
\end{enumerate}

\paragraph*{Related games.} Litinski's \emph{A Game of Surface Codes}~\cite{litinski2019game,pennylane_game} turns lattice surgery into a board game with explicit space and time costs, and was the template for our spacetime board. \emph{Quantum Poker}~\cite{fuchs2019poker} uses a game to benchmark error mitigation, Decodoku and its genetic-algorithm solver~\cite{wootton2016minimal,genetic_decodoku} turn decoding into a puzzle, and Ref.~\cite{guerrero2025nash} uses game theory to search for codes. To our knowledge no existing game is built on error detection with spacetime codes, nor on the interplay between check placement and non-Clifford doping.

%% file: sections/background.tex
\section{Background}
\label{sec:background}

\subsection{Stabilizers, syndromes and post-selection}
An $n$-qubit Clifford circuit $U$ applied to $\ket{0}^{\otimes n}$ prepares a stabilizer state $\ket{\psi}$, the unique joint $+1$ eigenstate of the abelian group $\mathcal{S}=\{UZ_AU^\dagger : A\subseteq[n]\}$ of $2^n$ Pauli operators~\cite{gottesman1997stabilizer}. Such states can be simulated efficiently~\cite{aaronson2004improved}, which is why Clifford circuits are the natural playground for error detection: we know, at every point in the circuit, which Pauli operators should have a definite value. Measuring one of them and finding the wrong value reveals that an error occurred. A collection of such measurements produces a syndrome bit string $s$; error detection keeps a shot only if $s=0$. If $\rho$ is the noisy state of the data and the ancillas and $\Pi_0$ the projector onto a trivial syndrome, the kept fraction and the post-selected state are
\begin{equation}
  \kappa=\mathrm{Tr}(\Pi_0\rho),\qquad \sigma=\frac{\mathrm{Tr}_{\mathrm{anc}}(\Pi_0\rho\Pi_0)}{\kappa}.
  \label{eq:postselect}
\end{equation}
Detection improves the fidelity of $\sigma$ with the ideal state at the price of a smaller $\kappa$. A post-selected fidelity is meaningless without its kept fraction, and we never report one without the other.

\subsection{Spacetime Pauli checks}
A spacetime check~\cite{martiel2025spacetime} uses one ancilla, prepared in $\ket{+}$, which applies controlled Pauli gates $P_1,\dots,P_m\in\{X,Y,Z\}$ to a single target qubit at wires (moments) $w_1,\dots,w_m$ of the payload, after which the ancilla is rotated back with a Hadamard and measured. Let $B(P,w)$ denote $P$ propagated backwards from wire $w$ to the start of the circuit. The check is \emph{valid} if $\prod_iB(P_i,w_i)$ belongs to the stabilizer group of the input, up to a Pauli $Z_S$ on output qubits that is read out classically: the syndrome bit is then the parity of the ancilla outcome and of the output bits in the support $S$. Equivalently, one can follow the \emph{flow} of the check forward through the circuit. Between two consecutive controlled gates the ancilla is entangled with a Pauli operator $C(t)$ on the data, and a fault $E$ at spacetime location $t$ flips the syndrome exactly when $E$ anticommutes with $C(t)$. Wires where $C(t)$ is the identity are not monitored at all; wires where $C(t)$ acts as $Z$ on a qubit catch $X$ and $Y$ errors there, and so on. The art is to find checks that are valid, use few entangling gates, and cover the faults that matter. \texttt{qiskit-paulice}~\cite{paulice} searches for such checks one target at a time and scores candidate check sets by $\gamma$, the sampling overhead of inverting the noise channel left uncovered by the checks. A value $\gamma=1$ means all modelled noise is detectable.

Coherent Pauli checks~\cite{vandenberg2023cpc} measure a single, generally high-weight operator at one time. On hardware with limited connectivity this costs many SWAP gates. Spreading the check over time, as above, needs only couplings between the ancilla and one neighbouring target, which is what makes the scheme low overhead~\cite{martiel2025spacetime,ibm_tutorial_spacetime}.

\subsection{Magic, and why $T$ gates are special}
Clifford circuits are classically simulable, so a useful quantum computation needs a non-Clifford resource, conventionally the $T=\mathrm{diag}(1,e^{i\pi/4})$ gate~\cite{bravyi2005magic}. \texttt{add\_pauli\_checks} only accepts Clifford circuits, so $T$ gates must be added after the checks are found. Because $T$ is diagonal, $TZT^\dagger=Z$, while $TXT^\dagger=(X+Y)/\sqrt2$ is not a Pauli operator. This gives the rule stated in the challenge and used by Martiel \emph{et al.}~\cite{martiel2026sampling}:

\begin{lemma}[Doping rule]
\label{lem:doping}
Let a set of checks be valid for a Clifford circuit, and insert a $T$ gate on qubit $q$ at spacetime location $t$. If, for every check, the flow operator $C(t)$ acts on $q$ as $I$ or $Z$, all checks remain valid. If some check has $C(t)$ acting on $q$ as $X$ or $Y$, that check is in general no longer valid.
\end{lemma}
\begin{proof}
If $C(t)$ acts as $I$ or $Z$ on $q$ then $T C(t) T^\dagger=C(t)$, so the flow of the check after $t$, and therefore the deterministic ideal value of its syndrome, is unchanged. If $C(t)$ acts as $X$ or $Y$, then $TC(t)T^\dagger$ is a superposition of two Pauli operators that differ on $q$, so the flow does not close onto a stabilizer and the ancilla outcome becomes random on the ideal circuit.
\end{proof}

A convenient way to find the safe locations is to ask where a single $Z$ error would be invisible to every check: those are exactly the locations where every flow acts as $I$ or $Z$. \texttt{qiskit-paulice} reports them through \texttt{CheckedCircuit.uncovered\_paulis}. We call them \emph{Z-safe sites}. A second hardware fact matters here. On IBM processors $R_Z$ rotations, and therefore $T$ gates, are implemented as virtual frame changes and add no error~\cite{mckay2017virtualz}. The $T$ count of a circuit thus costs nothing in noise, but, by Lemma~\ref{lem:doping}, it can cost a great deal in detection power. This asymmetry drives the central trade-off of our game.

\subsection{Toward fault tolerance}
The guiding question of the challenge is what fault tolerance is and what it needs~\cite{ibm_blog_ft}. A fault-tolerant computer must keep the error rate of every logical operation below a target while the computation runs, which requires codes with growing distance, repeated syndrome extraction, fast decoding, and a supply of high-fidelity magic states~\cite{fowler2012surface,bravyi2024high,google2025below,bravyi2005magic}. Spacetime detection gives up on the first three and asks a more modest question: how much of today's noise can a handful of ancillas catch in one shot? It is a stepping stone that already exposes two of the hard parts of fault tolerance, namely that protection costs qubits, gates and time, and that non-Clifford resources do not sit comfortably inside a stabilizer code.

%% file: sections/game.tex
\section{The game}
\label{sec:game}

\subsection{Board and moves}
\begin{figure*}[t]
\centering
\includegraphics[width=\textwidth]{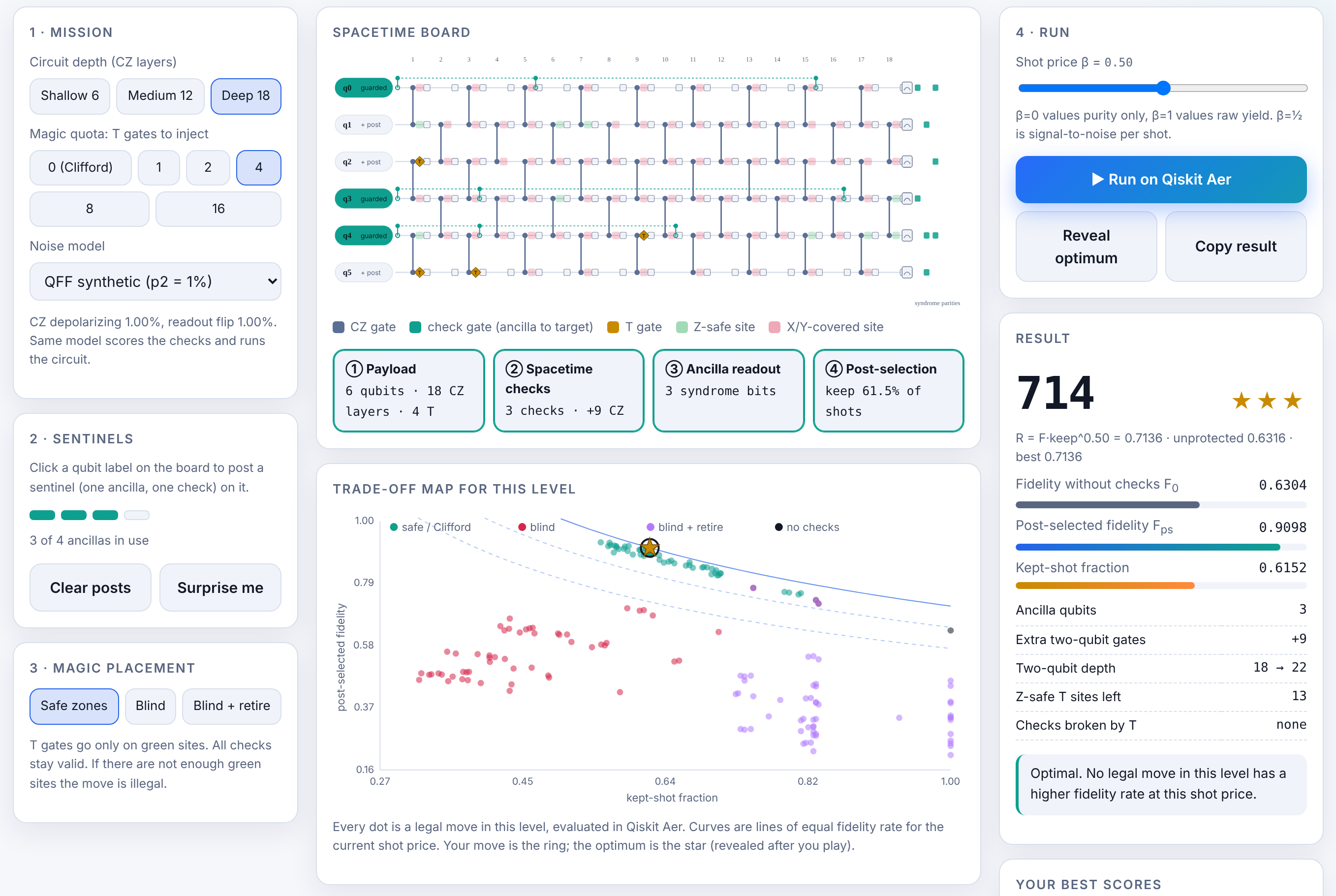}
\caption{The web game during a deep level (18 CZ layers, quota of 4 $T$ gates). Sentinels are posted on three qubits; teal probes show where each check touches its target, green and red squares mark Z-safe and covered $T$ sites, gold diamonds are the placed $T$ gates, and the right panel shows the Qiskit result of the move. The lower panel plots every legal move of the level.}
\label{fig:board}
\end{figure*}

The board (Fig.~\ref{fig:board}) is the spacetime diagram of the payload: six horizontal wires for the data qubits and one column per layer of CZ gates. A level fixes the depth (6, 12 or 18 layers), the \emph{magic quota} (0, 1, 2, 4, 8 or 16 $T$ gates) and the noise model. The player then makes three decisions, which are precisely the decisions of steps 3 and 6 of the challenge workflow:

\begin{description}[leftmargin=1.2em,itemsep=2pt]
  \item[Post sentinels.] Clicking a qubit label posts a sentinel there: one ancilla that will carry one spacetime check targeting that qubit. The ancilla budget is four. \texttt{add\_pauli\_checks} then decides the Pauli types and the moments at which the check touches its target, and the board draws them as teal probes.
  \item[Inject magic.] The quota of $T$ gates must be placed. After sentinels are posted, every candidate $T$ site is painted green (Z-safe for all checks) or red (covered by an $X$ or $Y$ part of some check flow).
  \item[Choose a placement policy.] \emph{Safe zones} puts the $T$ gates on green sites only, following a fixed random order; if fewer green sites exist than the quota, the move is illegal. \emph{Blind} places them on the first sites of the same order, ignoring the checks. \emph{Blind and retire} uses blind placement but stops post-selecting on any check that the $T$ gates have broken.
\end{description}
\noindent Pressing \emph{Run} reveals the four stages of the pipeline required by the challenge (payload, spacetime checks, ancilla readout, post-selection filter) together with the fidelity without checks $F_0$, the post-selected fidelity $F_{\mathrm{ps}}$, the kept fraction $\kappa$, the cost of the checks (ancillas, extra two-qubit gates, two-qubit depth) and which checks, if any, were broken.

\subsection{Score}
\label{sec:score}
The score is $1000\,R_\beta$ with the \emph{fidelity rate}
\begin{equation}
  R_\beta = F_{\mathrm{ps}}\;\kappa^{\beta}, \qquad 0\le\beta\le1 .
  \label{eq:rate}
\end{equation}
The exponent $\beta$ is the price of a discarded shot and is a slider in the game. Its default value has an operational meaning. Suppose the error left after post-selection damps a traceless observable $O$ by the factor $F_{\mathrm{ps}}$, as a depolarizing residual channel would, and suppose we run $N$ shots and keep $\kappa N$. The estimate of $\langle O\rangle$ obtained by rescaling the post-selected mean by $1/F_{\mathrm{ps}}$ has a standard error of roughly $\sigma_O/(F_{\mathrm{ps}}\sqrt{\kappa N})$, so reaching a precision $\varepsilon$ requires
\begin{equation}
  N \approx \frac{\sigma_O^2}{\varepsilon^2}\,\frac{1}{F_{\mathrm{ps}}^2\,\kappa}=\frac{\sigma_O^2}{\varepsilon^2}\,R_{1/2}^{-2}.
\end{equation}
Hence $R_{1/2}$ is the signal-to-noise ratio per square root of a raw shot, and $R_{1/2}^{-2}$ is the total sampling overhead of detection followed by a rescaling step, which is the detection-plus-mitigation workflow suggested in the \texttt{qiskit-paulice} documentation~\cite{paulice}. Without checks the same argument gives $R=F_0$, so the ratio $G=R_{1/2}/F_0$ tells the player whether the sentinels beat doing nothing. The two extremes are also meaningful: $\beta=0$ rewards purity regardless of cost and $\beta=1$ rewards the yield $F_{\mathrm{ps}}\kappa$, which, for valid checks, can never exceed the fidelity of the unpost-selected checked circuit. A move earns one star if it beats the unprotected circuit, two if it is within 2\% of the best legal move of the level, and three if it is the best.

\subsection{Rule to physics}
Table~\ref{tab:map} lists every rule and the physics it stands for. Two design choices deserve comment. We let \texttt{qiskit-paulice} choose the internal structure of each check rather than letting the player draw controlled-Pauli gates by hand, because a hand-drawn check is almost always invalid and the resulting game would be about bookkeeping rather than about trade-offs. The spatial choice (which qubits carry an ancilla) and the interaction with the $T$ gates remain with the player, and those are the choices that matter. Second, the game keeps the bad options. Blind placement and retiring broken checks are legal because players should discover for themselves, with real numbers, why they are bad.

\begin{table*}[t]
\caption{Game rules and the physics they represent. All quantities are computed with Qiskit; none is hand-set.}
\label{tab:map}
\begin{ruledtabular}
\begin{tabular}{p{0.21\textwidth}p{0.47\textwidth}p{0.26\textwidth}}
Game element & Physics & Computed by \\
\hline
Qubit wire, CZ column & Data qubit, layer of a brickwork Clifford payload & \texttt{QuantumCircuit} \\
Sentinel post (budget 4) & One ancilla and one spacetime check targeting that data qubit & \texttt{add\_pauli\_checks} (\texttt{gamma}, \texttt{windowed}) \\
Teal probes on a wire & Moments at which the check applies a controlled Pauli to its target & checked circuit \\
Syndrome parities at readout & Output bits XORed with the ancilla bit to form the syndrome & \texttt{check\_support} \\
Green / red site & Every check flow acts as $I/Z$ there (T is safe) / some flow acts as $X/Y$ & \texttt{uncovered\_paulis} \\
Magic quota & Number of non-Clifford $T$ gates in the payload & $T$ inserted after check search \\
Safe zones / Blind / Retire & Doping policy; a $T$ on a red site breaks a check (Lemma~\ref{lem:doping}) & noiseless \texttt{Statevector} test \\
Run & Noisy execution with ancilla readout and post-selection & Aer density matrix, exact \\
$F_0$, $F_{\mathrm{ps}}$, $\kappa$ & Fidelity with the ideal state, without and with post-selection; kept fraction & stabilizer or statevector reference \\
Score $1000R_\beta$, stars & Fidelity rate, Eq.~\eqref{eq:rate}; comparison with every legal move & \texttt{sentinels.scoring} \\
\end{tabular}
\end{ruledtabular}
\end{table*}

\subsection{Implementations}
The web version, published with GitHub Pages at \url{https://sumitchongder.github.io/spacetime-sentinels/}, is a static page that plays back the Qiskit results for every legal move of every level, so it runs instantly on any device. Each level also shows the full trade-off map of the level, every legal move plotted as $F_{\mathrm{ps}}$ against $\kappa$ with curves of constant $R_\beta$, and reveals the optimum after the player has committed. The terminal version, \texttt{sentinels play}, computes the chosen move live with \texttt{qiskit-paulice} and Qiskit Aer, and a notebook walks through the six steps of the challenge.

%% file: sections/methods.tex
\section{Methods}
\label{sec:methods}

The numbered paragraphs below follow the six required steps of the challenge.

\paragraph*{Step 1: payloads.} The Clifford payload is the one-dimensional brickwork used in the IBM tutorial~\cite{ibm_tutorial_spacetime}: a layer of Hadamards on $n=6$ qubits followed by $d$ layers, each made of CZ gates on alternating nearest-neighbour pairs and a random choice of $\sqrt{X}$, $S$, $\sqrt{X}$ on every qubit. We use $d=6$, $12$ and $18$, giving 15, 30 and 45 CZ gates, a width of six qubits and a two-qubit depth equal to $d$ (Table~\ref{tab:payloads}). Every CZ gate defines two candidate $T$ sites, one on each of its qubits immediately after the gate, so the three payloads offer 30, 60 and 90 sites. The $T$-doped family places $n_T\in\{0,1,2,4,8,16\}$ $T$ gates on these sites. The game uses payload seed 7; the study in Sec.~\ref{sec:results} repeats every experiment for five seeds (7, 11, 13, 17, 19).

\begin{table}[b]
\caption{Payloads. The two-qubit depth of the checked circuit depends on the checks and is reported with every result.}
\label{tab:payloads}
\begin{ruledtabular}
\begin{tabular}{lccc}
 & shallow & medium & deep \\
\hline
CZ layers $d$ & 6 & 12 & 18 \\
Width (data qubits) & 6 & 6 & 6 \\
CZ gates & 15 & 30 & 45 \\
Candidate $T$ sites & 30 & 60 & 90 \\
$F_0$, Clifford, QFF noise & \FzeroShallow & \FzeroMedium & \FzeroDeep \\
\end{tabular}
\end{ruledtabular}
\end{table}

\paragraph*{Step 2: reference and fidelity measure.} For the Clifford payload the ideal output $\ket{\psi}$ is obtained by stabilizer simulation (\texttt{StabilizerState}); for $T$-doped payloads by statevector simulation (\texttt{Statevector}), which is exact and cheap at six qubits. We use the state fidelity of the post-selected data register,
\begin{equation}
  F_{\mathrm{ps}}=\bra{\psi}\sigma\ket{\psi},
\end{equation}
because it is sensitive to every error, including phase errors that any computational-basis distance misses. This matters here: the brickwork payloads have nearly flat output distributions, and for the deep Clifford payload of the game the classical (Hellinger) fidelity of the noisy output is \HzeroDeep{} while its state fidelity is only \FzeroDeep. For Clifford states we evaluate $F$ through the identity $F=2^{-n}\sum_{G\in\mathcal S}\mathrm{Tr}(\sigma G)$, which is what direct fidelity estimation measures~\cite{flammia2011direct,dasilva2011practical}, and check that it agrees with the statevector overlap to machine precision.

One technical point needs care. A syndrome bit is the parity of an ancilla outcome and of some output bits. In an experiment the parity is computed classically after all qubits are measured, so the post-selected object is a distribution, not a state. In simulation we obtain the corresponding state by computing the parity coherently: after the readout channel we apply noiseless CNOT gates from the output qubits in the support $S$ onto the ancilla, so that the ancilla holds the syndrome. These gates act after all noise and are diagonal on the data, so every computational-basis statistic, including $\kappa$ and the post-selected distribution, is identical to the experiment. The data state $\sigma$ of Eq.~\eqref{eq:postselect} is then well defined, and on an ideal run with valid checks $\sigma=\ket\psi\bra\psi$ exactly. Simply tracing out the ancillas without this step would dephase the data in the eigenbasis of $Z_S$ and make even a perfect run look unfaithful.

\paragraph*{Step 3: checks.} Checks are found with \texttt{qiskit-paulice} 0.2.1, \texttt{add\_pauli\_checks} with \texttt{cost="gamma"}, \texttt{method="windowed"} and \texttt{seed=1}, the deterministic setting recommended by its documentation. The payload, terminated by measurements, is passed with the chosen target qubits in increasing order; each target receives its own ancilla. The function returns one circuit per number of checks, which we use in Sec.~\ref{sec:res-k}; for the game each target set is a separate call. The checked circuit is re-synthesised by the add-on and may reorder commuting CZ gates, so we match payload gates by their position along each wire, and we obtain the unprotected reference circuit by deleting all ancilla gates, which leaves data gates that compose to the original payload (verified in the test suite). For $T$-doped payloads the add-on cannot be called. We use the checks found on the Clifford skeleton and apply Lemma~\ref{lem:doping}: a site is Z-safe if \texttt{uncovered\_paulis} lists a $Z$ error there. Independently of that list, we test every doped circuit by statevector simulation without noise, and declare a check broken if its syndrome is non-zero with probability above $10^{-9}$. All safe-zone moves pass this test with false-alarm rates that vanish to numerical precision.

\paragraph*{Step 4: noise.} The challenge suggests the synthetic model \texttt{NoiseModel(gate\_noise=0.01)}. Our main profile, \emph{QFF synthetic}, is exactly that for the check search, and the same channel for simulation: a two-qubit depolarizing channel with $p_2=0.01$ after every CZ, payload or check, and a symmetric readout flip with $p_{\mathrm{ro}}=0.01$ on every qubit. Single-qubit Cliffords and $T$ gates are noiseless, the latter because they are virtual on IBM hardware~\cite{mckay2017virtualz}. A second profile, \emph{FakeFez}, follows the tutorial and derives a uniform model from backend calibration data, \texttt{NoiseModel.from\_backend(FakeFez(), layout, uniform\_gate\_noise=True)}, giving $p_2=\PtwoFez$ and $p_{\mathrm{ro}}=\ProFez$. Idling noise is ignored in both, as it is in the current check picker. Using the same channel for picking and for simulating matches the tutorial, so the checks are tuned to the noise they face. If the organisers announce a different model, the whole pipeline is re-run with one command.

\paragraph*{Step 5: measurement.} For every configuration we run two exact Qiskit Aer density-matrix simulations: the checked circuit with its ancillas, which gives $\kappa$ and $F_{\mathrm{ps}}$, and the unprotected circuit with the same $T$ gates, which gives $F_0$. We record the ancilla count, the number of extra two-qubit gates and the two-qubit depth with and without checks, the $\gamma$ cost, the number of Z-safe sites, the broken checks and the classical fidelity. Exact simulation removes shot noise, so two moves that differ by a fraction of a percent can be ranked reliably. During testing we found that noiseless density-matrix runs disagreed with \texttt{Statevector} by up to $3\times10^{-2}$ when Aer's gate fusion was on; with fusion off the agreement is $10^{-16}$, so fusion is disabled throughout.

\paragraph*{Validation.} Exact numbers are only useful if they describe what an experiment would see. We therefore re-ran twenty configurations, covering both noise models, all three depths, Clifford and doped payloads and all three placement policies, through the Qiskit Aer \texttt{SamplerV2} primitive with 50{,}000 shots each, computing syndromes from the measured bits exactly as one would on hardware. The largest deviation of the kept fraction from the exact value is \ValKeepSigma{} standard errors and the largest deviation of the classical fidelity is \ValHdiff{} (Fig.~\ref{fig:validation}).

\paragraph*{Step 6: comparison and search.} The decision space of the game is small enough to enumerate: 57 target sets with at most four checks, five non-zero $T$ quotas with three policies each, three depths and two noise models, \NumMoves{} moves in total. We evaluated all of them. The research experiments add check-count sweeps with up to six checks (E1), $T$-count sweeps up to 16 $T$ gates (E2), magic capacity for all 64 target sets at four depths (E3), a sweep of the noise strength from $p_2=10^{-3}$ to $3\times10^{-2}$ (E5) and an ablation with noisy $T$ gates (E7), each over several payload seeds. Error bars show the standard deviation over seeds.

\paragraph*{Software and compute.} Qiskit 2.5.2, Qiskit Aer 0.17.2, \texttt{qiskit-paulice} 0.2.1, \texttt{qiskit-ibm-runtime} 0.50.0 (fake backends only) and Python 3.13. Every number in this report was produced on a two-core laptop-class machine; the full sweep of the game took about \GridMinutes{} minutes per noise model. The configuration-level parallelism is exposed through a process pool and through Slurm array scripts, and Aer can be switched to GPU execution for larger studies.

%% file: sections/results.tex
\section{Results}
\label{sec:results}

Unless stated otherwise, results use the QFF synthetic noise model and are averaged over five random payloads; error bars are standard deviations over payloads. Numbers quoted for ``the game'' refer to the single payload (seed 7) that the game uses.

\subsection{Validation}
\label{sec:res-val}
\begin{figure}[t]
\centering
\includegraphics[width=\columnwidth]{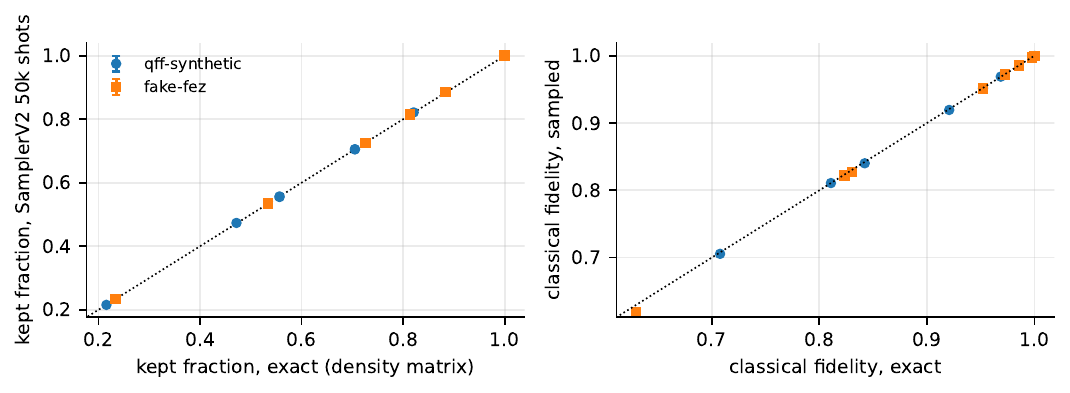}
\caption{Exact density-matrix numbers against 50{,}000-shot runs of the Aer \texttt{SamplerV2} primitive, for twenty configurations and both noise models. Error bars are two standard errors.}
\label{fig:validation}
\end{figure}

Figure~\ref{fig:validation} compares the exact evaluation with 50{,}000-shot \texttt{SamplerV2} runs for twenty configurations. The kept fractions agree within \ValKeepSigma{} standard errors in the worst case, and the classical fidelities of the post-selected distributions agree to within \ValHdiff. Noiseless runs confirm that every check found by \texttt{qiskit-paulice}, and every check after safe-zone doping, has a false-alarm rate that is numerically zero (below $10^{-9}$ in all \NumMoves{} moves), while each check broken by blind doping fires on perfect runs with probability between 0.15 and 0.69 (median 0.25). For Clifford payloads the stabilizer-group identity and the statevector overlap give the same $F_0$ to within \ValStabDiff.

\subsection{How many checks? A break-even depth}
\label{sec:res-k}
\begin{figure*}[t]
\centering
\includegraphics[width=0.9\textwidth]{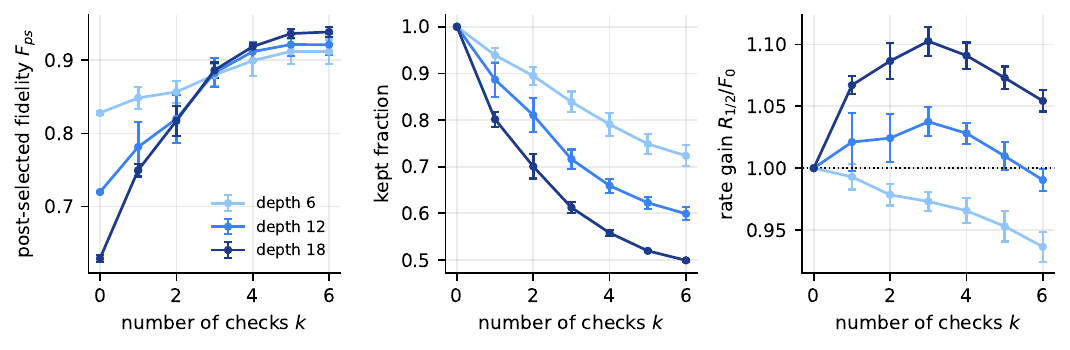}
\caption{Post-selected fidelity, kept fraction and fidelity-rate gain $R_{1/2}/F_0$ against the number of checks committed by \texttt{add\_pauli\_checks} (targets 0 to 5 in order), for three depths. Five payloads; error bars are standard deviations. The dotted line marks break-even.}
\label{fig:checks}
\end{figure*}

Figure~\ref{fig:checks} follows one call of \texttt{add\_pauli\_checks} that commits checks on qubits 0 to 5 one at a time. More checks always raise the post-selected fidelity and always lower the kept fraction, as expected. The interesting quantity is the gain in fidelity rate, $G=R_{1/2}/F_0$. At depth 18 the fidelity rises from \EoneFzeroDeep{} without checks to \EoneFpsDeep{} with the best number of checks ($k=\EoneKDeep$) while \EoneKeepDeep{} of the shots survive, a gain of \EoneGainDeep. At depth 12 the best gain is \EoneGainMedium{} ($k=\EoneKMedium$). At depth 6 no number of checks beats the unprotected circuit in this sweep: one check already gives $G=\EoneGainOneShallow$, and every further check lowers the rate. Choosing the targets carefully helps only a little; the best move in the game's shallow Clifford level, checks on \BestShallowTZeroTargets, gains a factor of \BestShallowTZeroGain{} (Table~\ref{tab:best}).

The reason is a budget argument. A check adds two to five CZ gates, each of which can fail with probability $p_2$, and it can only catch errors that occur on the wires its flow monitors. On a shallow circuit there are few errors to catch, so the check's own gates are a large fraction of the total noise, and most of what the extra checks catch are errors that the checks themselves introduced. As the circuit gets deeper the payload contributes more and more of the noise, and the same ancilla becomes a better investment. The game makes this visible: in the shallow levels the best possible score is barely above the score of posting no sentinel at all, and most moves score below it.

\subsection{$T$ count against check placement}
\label{sec:res-t}
\begin{figure*}[t]
\centering
\includegraphics[width=0.95\textwidth]{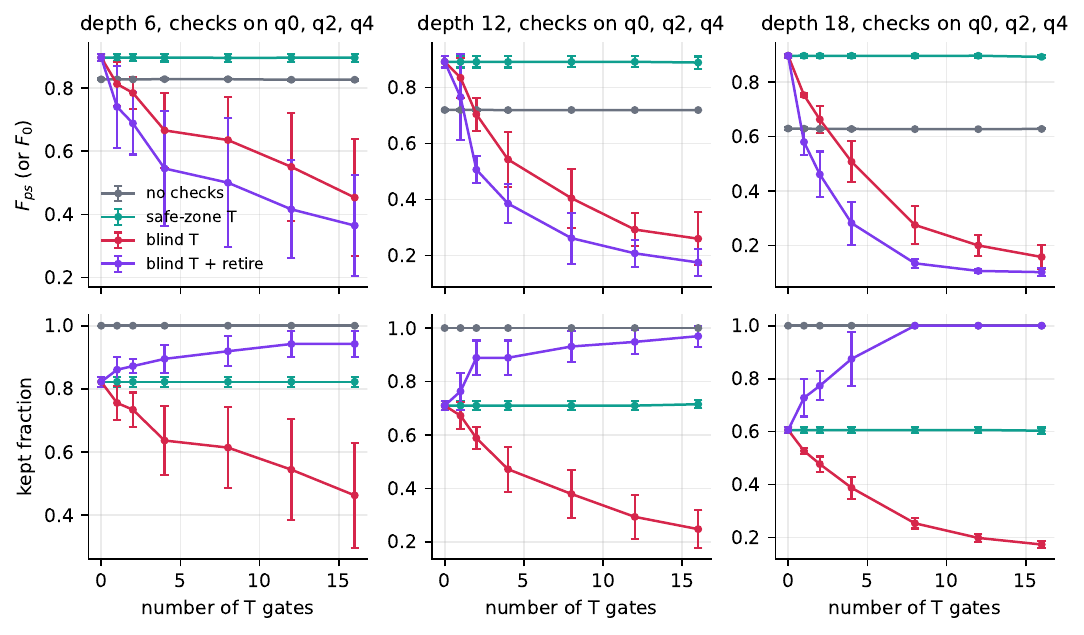}
\caption{$T$ count against check placement. Top: post-selected fidelity (or $F_0$ without checks); bottom: kept fraction; columns: depth 6, 12, 18. Checks on qubits 0, 2, 4. Safe-zone doping (teal) is flat; blind doping (red) breaks checks; retiring broken checks (purple) is worse than having no checks (grey).}
\label{fig:tcount}
\end{figure*}

Figure~\ref{fig:tcount} is the central result. A fixed set of three checks (on qubits 0, 2 and 4) protects payloads doped with up to 16 $T$ gates, placed with each of the three policies. Without checks, the $T$ gates barely change the fidelity: at depth 18 $F_0$ moves from \EtwoNoneFzeroTZeroDeep{} with no $T$ gate to \EtwoNoneFzeroTSixteenDeep{} with sixteen, because the $T$ gates themselves are noiseless and only slightly change how the existing errors propagate. With safe-zone doping, the post-selected fidelity and the kept fraction are flat as well: the checks see exactly the same faults whether or not the $T$ gates are present, which is the invariance exploited in Ref.~\cite{martiel2026sampling}. The largest drop in $F_{\mathrm{ps}}$ between the Clifford circuit and its 16-$T$ safe-zone version is \EtwoSafeDrift{}, the footprint of errors that were harmless before doping and are harmful after it.

Blind doping tells a different story. Each $T$ that lands on a wire where some check flow acts as $X$ or $Y$ breaks that check. With only four $T$ gates at depth 18, blind placement breaks on average \EtwoBrokenFourDeep{} of the three checks. A broken check fires on perfect runs, so post-selecting on it throws away good shots: at depth 18 with 16 $T$ gates, the kept fraction falls to \EtwoBlindKDeepTSixteen{} and the post-selected fidelity to \EtwoBlindFDeepTSixteen, against \EtwoSafeKDeepTSixteen{} and \EtwoSafeFDeepTSixteen{} with safe-zone doping. Post-selecting on a broken check is a projection onto one eigenspace of an operator that no longer stabilizes the ideal state, and that projection rotates the state away from the target.

The third policy is the one most players reach for: if a check is broken, stop listening to it. This is worse still. With 16 $T$ gates at depth 18 the fidelity of blind doping with retirement is \EtwoRetireFDeepTSixteen, below the \EtwoNoneFDeepTSixteen{} of the circuit with no checks at all. The ancilla of a retired check is still entangled with the data through a flow that no longer closes, so discarding its outcome amounts to measuring a non-stabilizer Pauli operator and forgetting the result. That is a dephasing channel on the data, and it destroys exactly the coherence that the $T$ gates were meant to create. A broken check cannot be switched off after the fact; it has to be removed from the circuit or never broken in the first place.

\subsection{Magic capacity: protection and magic compete}
\label{sec:res-cap}
\begin{figure*}[t]
\centering
\includegraphics[width=0.9\textwidth]{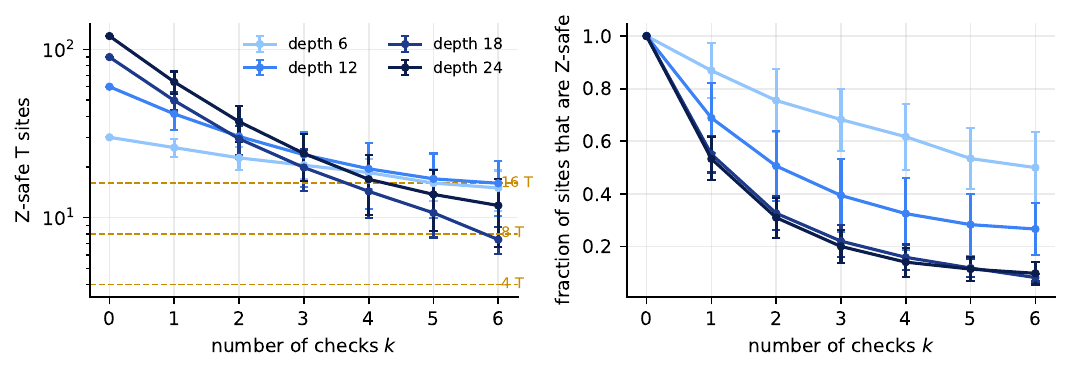}
\caption{Magic capacity. Left: number of Z-safe $T$ sites against the number of checks, averaged over all target sets of that size and five payloads (log scale; dashed lines mark $T$ demands of 4, 8 and 16). Right: the same as a fraction of all sites.}
\label{fig:capacity}
\end{figure*}

If safe-zone doping is free, why not always use it? Because the safe zones shrink as checks are added. Figure~\ref{fig:capacity} counts the Z-safe sites for every one of the 64 target sets at four depths. At depth 18 all \CapTotalDeep{} sites are safe without checks; with two checks \CapDeepTwo{} remain on average, with four \CapDeepFour{}, and with six only \CapDeepSix. A check is useful precisely because its flow acts as $X$ or $Y$ on many wires, where it catches $Z$ errors, and those are exactly the wires where a $T$ gate would break it. Protection against phase errors and room for magic are two uses of the same spacetime, and one can only be bought at the expense of the other.

In the game this shows up as illegal moves. Across all levels \NumIllegal{} moves are illegal because there are fewer safe sites than $T$ gates to place, all of them in the levels with 16 $T$ gates. The optimum moves accordingly. In the game's deep level, the best move with up to eight $T$ gates uses three checks (\BestDeepTEightTargets), while with sixteen $T$ gates the best move drops to \BestDeepTSixteenTargets: one fewer check, to free enough safe sites for the magic. The gain falls from \BestDeepTEightGain{} to \BestDeepTSixteenGain. This is the non-obvious trade-off asked for by the challenge in its plainest form: the right number of checks depends on how much magic the circuit carries.

\subsection{The best trade-off, and why it wins}
\label{sec:res-best}
\begin{figure}[t]
\centering
\includegraphics[width=\columnwidth]{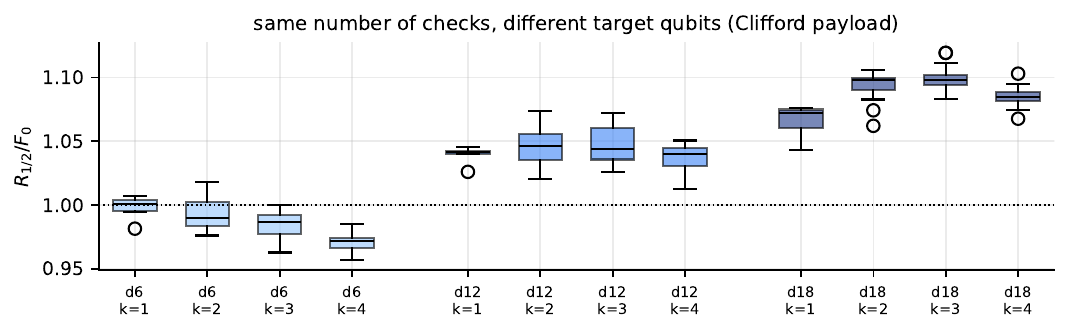}
\caption{Check location. Fidelity-rate gain of every target set with $k$ checks on the Clifford payloads of the game.}
\label{fig:location}
\end{figure}

Table~\ref{tab:best} lists the best move of every level of the game under the default shot price $\beta=1/2$. Three features stand out. First, the winning moves never use blind doping when the quota is four or more: as soon as a check can be broken, protecting it pays more than any extra coverage. Second, the winning moves use two or three checks, never four. The fourth check adds gates for little extra coverage, the same diminishing return seen in the $\gamma$ cost and in Fig.~\ref{fig:checks}. Third, which qubits are guarded matters almost as much as how many. With two checks on the deep payload the gain ranges from \LocMinDeep{} to \LocMaxDeep{} depending on the targets (Fig.~\ref{fig:location}). The $\gamma$ cost used by \texttt{qiskit-paulice} ranks checks for a fixed target, but it does not choose the targets; the player does, and a poor choice of targets can waste most of the benefit.

\begin{figure*}[t]
\centering
\includegraphics[width=0.9\textwidth]{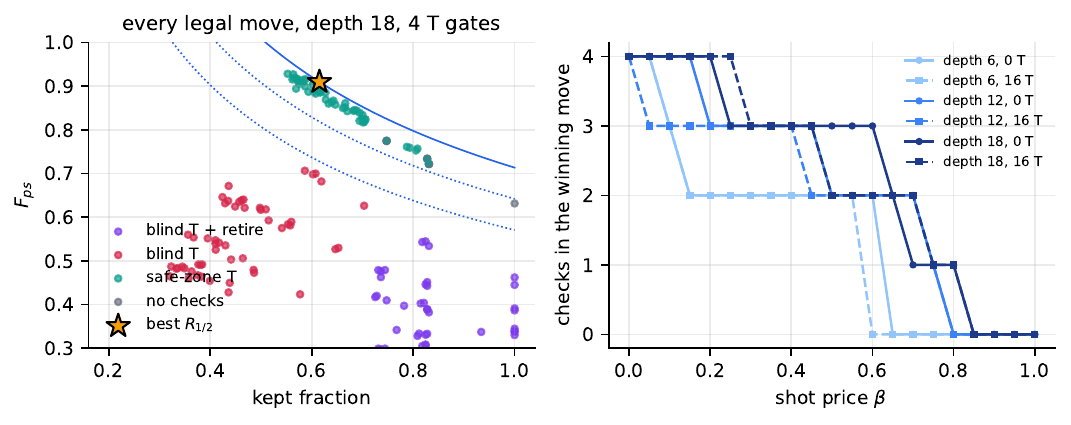}
\caption{Left: every legal move of the game's deep level with four $T$ gates. Lines are curves of constant $R_{1/2}$ through the best move (star) and at 90\% and 80\% of it. Right: number of checks in the winning move against the shot price $\beta$, for Clifford payloads (solid) and 16 $T$ gates (dashed).}
\label{fig:pareto}
\end{figure*}

Why does the deep, four-$T$ winner (\BestDeepTFourTargets, safe zones) win? It raises the fidelity from \BestDeepTFourFzero{} to \BestDeepTFourFps{} while keeping \BestDeepTFourKeep{} of the shots. Its three checks sit on qubits whose flows overlap little, so together they cover a large share of the two-qubit faults, and the rejection rate is close to the probability that a fault occurred at all. Adding a fourth check would catch fewer new faults than its own gates create, and any extra check would also remove safe sites. The move sits on the Pareto front of the level (Fig.~\ref{fig:pareto}), at the point where the curve of constant $R_{1/2}$ touches it.

The shot price changes the answer (Fig.~\ref{fig:pareto}, right). At $\beta=0$, where only purity counts, the best move uses all four checks at every depth; at $\beta=1$, where every discarded shot counts fully, it uses none. The default $\beta=1/2$ is the only one of the three with an operational meaning (Sec.~\ref{sec:score}), and it is the one in which spacetime detection is worth something on deep circuits and nothing on shallow ones.

\begin{table}[t]
\caption{Best move of each level of the game (payload seed 7, QFF synthetic noise, $\beta=1/2$). $G=R_{1/2}/F_0$.}
\label{tab:best}
\begin{ruledtabular}
\begin{tabular}{cclccc}
$d$ & $n_T$ & checks, placement & $F_0$ & $F_{\mathrm{ps}}$ ($\kappa$) & $G$ \\
\hline
\BestTableRows
\end{tabular}
\end{ruledtabular}
\end{table}

\subsection{Noise strength, noise model and noisy $T$ gates}
\label{sec:res-noise}
\begin{figure}[t]
\centering
\includegraphics[width=\columnwidth]{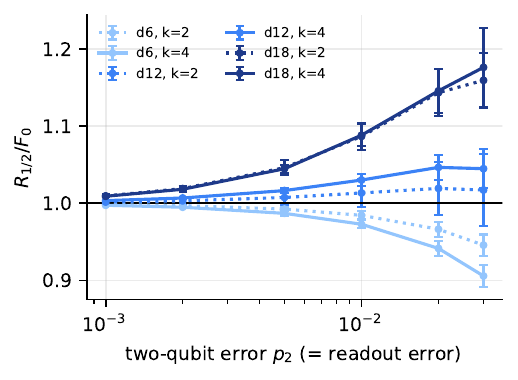}
\caption{Fidelity-rate gain against the two-qubit (and readout) error rate, for two and four checks at three depths. Checks are re-picked at every noise level.}
\label{fig:noise}
\end{figure}

Figure~\ref{fig:noise} sweeps the two-qubit and readout error together from $10^{-3}$ to $3\times10^{-2}$, re-picking the checks at every point. With four checks at depth 18 the gain grows from \EfiveGainDeepLow{} at $p_2=10^{-3}$ to \EfiveGainDeepHigh{} at $p_2=3\times10^{-2}$. At depth 6 it stays below one at every noise level and even falls, from \EfiveGainShallowLow{} to \EfiveGainShallowHigh, because the checks' own gates get noisier along with everything else. Detection is most useful exactly when the circuit is too noisy to run well without it, and only when the circuit is deep enough for the payload to dominate the noise budget.

\begin{figure}[t]
\centering
\includegraphics[width=\columnwidth]{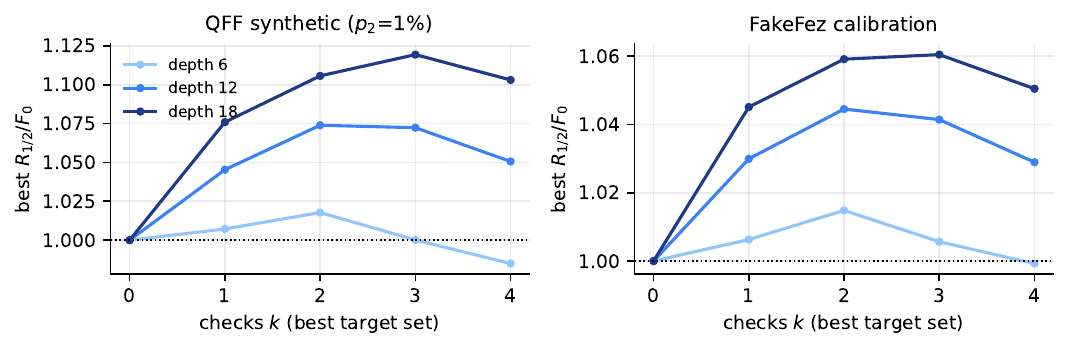}
\caption{Best gain per number of checks for the two noise models (Clifford payloads of the game).}
\label{fig:profiles}
\end{figure}

\begin{figure}[t]
\centering
\includegraphics[width=0.8\columnwidth]{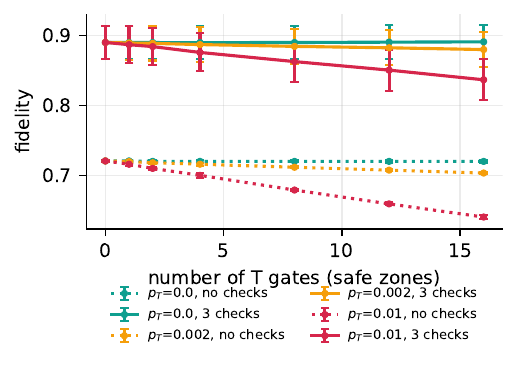}
\caption{Ablation with noisy $T$ gates (depth 12, safe-zone doping). Dotted: no checks; solid: three checks.}
\label{fig:tnoise}
\end{figure}

With the noise model derived from FakeFez calibration data, whose two-qubit error is about half that of the synthetic model, the same picture holds at smaller scale (Fig.~\ref{fig:profiles}): the best gain is \FezGainDeep{} at depth 18 with $k=\FezKDeep$ and \FezGainShallow{} at depth 6. Finally, if $T$ gates were not virtual but carried a depolarizing error of 1\%, the $T$ count would also lower the raw fidelity, from \EsevenFzeroHighTZero{} to \EsevenFzeroHighTSixteen{} at depth 12 (Fig.~\ref{fig:tnoise}). A $T$ gate sits on a Z-safe site, where every check flow acts as $I$ or $Z$, so the $Z$ part of its error is invisible to all checks and, wherever no flow passes, so is the rest. The post-selected fidelity therefore falls as well, from \EsevenFpsHighTZero{} to \EsevenFpsHighTSixteen. Safe sites are safe for the checks, not for the $T$ gates.

%% file: sections/discussion.tex
\section{Discussion}
\label{sec:discussion}

\subsection{What the game teaches}
The challenge lists four learning outcomes. The first, telling detection, mitigation and correction apart, is built into the rules: nothing in the game repairs a state, the only lever is which shots to keep, and the score itself (Sec.~\ref{sec:score}) is the overhead of detection followed by a mitigation-style rescaling, so the player sees where one ends and the other begins. The second, stabilizers and syndromes, appears in the probes and the syndrome parities drawn on the board and, more usefully, in the consequences of breaking them. The third, what makes practical quantum computing hard, is the content of our three findings: protection is made of the same noisy gates it protects against, it only pays above a break-even depth, and it collides with the non-Clifford resources that make a computation worth doing. The fourth, building on existing methods, is what the game asks of the player when choosing between safe and blind doping, which is the construction behind a recent fidelity-certified sampling experiment~\cite{martiel2026sampling}.

We find it useful that the bad options remain legal. A student who retires a broken check expects to recover the unprotected fidelity and instead lands below it. That surprise is the quickest route we know to the idea that an ancilla is not a passive observer: once it has interacted with the data, ignoring it is a choice with physical consequences.

\subsection{Relation to fault tolerance}
The guiding question asks what fault tolerance is and what it needs. Our results give a small-scale view of two of its requirements. The break-even depth is the detection analogue of a threshold: protection pays only when the errors it can catch outnumber the errors it introduces. The magic capacity is a toy version of a central difficulty of fault-tolerant architectures, where codes that protect Clifford operations transversally cannot do the same for $T$ gates, which must instead be supplied through costly magic-state factories~\cite{bravyi2005magic}. Here the conflict appears in its simplest form, as competition for wires, but the lesson carries over: the non-Clifford part of a computation does not fit comfortably inside a stabilizer code, and the cost of fitting it has to be planned for.

\subsection{Limitations}
We list the limitations we are aware of, because they bound what the numbers can be used for.
\begin{itemize}[leftmargin=*,itemsep=2pt]
  \item \textbf{Size.} Six data qubits and at most four checks per move (six in the check-count study) keep every move exactly simulable. The qualitative trends were stable across five payloads and two noise models, but the absolute gains are specific to this size; the tutorial and Ref.~\cite{martiel2026sampling} work at 12 to 76 qubits.
  \item \textbf{Noise model.} Uniform depolarizing gate noise and symmetric readout flips, with no idling noise, crosstalk or leakage. Idling noise is also ignored by the current check picker, so the checks are optimal only for the model they were picked with.
  \item \textbf{Coherent parity.} The parity of output bits is computed coherently in simulation to define a post-selected state. It reproduces all computational-basis statistics of the experiment, but the state fidelity it defines is not directly measurable on hardware without per-stabilizer checks as in the tutorial~\cite{ibm_tutorial_spacetime}.
  \item \textbf{Scope of the player's choices.} The player chooses where checks go and how $T$ gates are placed, not the internal structure of each check, which \texttt{qiskit-paulice} decides. A version in which players draw their own controlled-Pauli gates would teach more about validity and less about trade-offs.
  \item \textbf{Playback.} The web version plays back results computed in advance with Qiskit for every legal move. This is a deliberate choice that lets the game run on any device; the terminal version computes moves live.
  \item \textbf{No hardware run.} Everything runs on Qiskit Aer, as the challenge allows. The pipeline accepts any backend noise model and the shot-based path uses the same primitives interface as hardware execution.
\end{itemize}

\subsection{Future work}
Three extensions follow directly. A \emph{check-aware doping} optimiser could choose $T$ positions that keep the most coverage for a given magic budget, turning the capacity curve of Fig.~\ref{fig:capacity} into a design tool. A multiplayer mode in which one player places checks and another places $T$ gates would make the competition between protection and magic explicit, in the spirit of the game-theoretic approach of Ref.~\cite{guerrero2025nash}. Finally, the same scoring can be applied on hardware with \texttt{NoiseLearner}-based noise models and larger payloads, where the absolute gains reported by Ref.~\cite{martiel2025spacetime} are much larger than ours.

\section{Conclusion}
\label{sec:conclusion}
Spacetime Sentinels turns low-overhead error detection with spacetime codes into a game whose every rule is a decision of the real workflow and whose every score is a Qiskit result. Playing it, and the study behind it, show three things that are easy to state and hard to guess: spacetime checks only pay above a break-even depth; $T$ gates cost nothing in noise but compete with checks for the same spacetime, so the right number of checks depends on the amount of magic; and a check broken by a $T$ gate is worse than no check, even when ignored. The code, data, notebook and game are open source, and every figure in this paper can be regenerated with one command.

\begin{acknowledgments}
We thank the organisers of the Qiskit Fall Fest 2026 at IIT Jodhpur, Wenyu Sun for the challenge prompt, and IBM Quantum for \texttt{qiskit-paulice}, Qiskit and Qiskit Aer, and for the tutorials and Qiskit Global Summer School material~\cite{qgss2025lab4,qgss2026lab3} that the challenge builds on.
\end{acknowledgments}

\section*{Author contributions}
S.\,T.\,C. and A.\,K. contributed equally. Both authors designed the game, developed the software, ran the experiments, analysed the results and wrote the report.

\section*{Use of AI tools}
In line with the hackathon rules, we disclose that an AI assistant (Claude, by Anthropic) was used during this project to help write and refactor code, generate plotting and documentation boilerplate, and edit the wording of the report. The authors chose the problem framing and game design, verified the physical arguments, ran and inspected all experiments, and checked every number reported here against the generated data. All responsibility for the content rests with the authors.

\section*{Code and data availability}
All source code, raw and processed data (every evaluated move, as JSON and CSV), the scripts that regenerate every figure and number in this paper, a reproducible notebook, and the report and slides are openly available at \url{https://github.com/Sumitchongder/spacetime-sentinels} under the Apache 2.0 licence. The game can be played in any web browser at \url{https://sumitchongder.github.io/spacetime-sentinels/}. Open-source software used: Qiskit, Qiskit Aer, \texttt{qiskit-paulice} and \texttt{qiskit-ibm-runtime} (Apache 2.0), NumPy and Matplotlib.